\documentclass[twocolumn,aps,superscriptaddress,showpacs,nofootinbib,floatfix]{revtex4}
\usepackage{epsfig,bm,feynmf}
\usepackage{graphics}
\usepackage{amsmath}
\begin{document}

%%%%%%%%%%%%%%%%%%%%% Title %%%%%%%%%%%%%%%%%%%%%%

\title{Light nuclei production in Pb+Pb collisions at $\sqrt{s_{NN}}=2.76$ TeV}

%%%%%%%%%%%%%%%%%%%% Authors %%%%%%%%%%%%%%%%%%%%%

\author{Lilin Zhu}\email{zhulilin@scu.edu.cn}
\affiliation{Department of Physics, Sichuan University, Chengdu 610064, China}
\author{Hua Zheng}\email{zheng@lns.infn.it}
\affiliation{Laboratori Nazionali del Sud, INFN, via Santa Sofia, 62, 95123 Catania, Italy}
\author{Che Ming Ko}\email{ko@comp.tamu.edu}
\affiliation{Cyclotron Institute and Department of Physics and Astronomy, Texas A$\&$M University, College Station, TX 77843, USA}
\author{Yifeng Sun}\email{sunyfphy@physics.tamu.edu}
\affiliation{Cyclotron Institute and Department of Physics and Astronomy, Texas A$\&$M University, College Station, TX 77843, USA}

%%%%%%%%%%%%%%%%%%%% Abstract %%%%%%%%%%%%%%%%%%%%%

\begin{abstract}
Using the coalescence model based on the phase-space distribution of nucleons from an extended blast-wave model that includes the space-momentum correlation of high momentum nucleons, we study the transverse momentum spectra and elliptic flows of deuteron and helium-3 in Pb+Pb collisions at the energy $\sqrt{s_{NN}}=2.76$ TeV.  We find that the measured elliptic flows of deuteron can be satisfactorily described if nucleons of large transverse momenta are more spread in space when their momenta are more aligned along the reaction plane.
\end{abstract}

\maketitle

\section{introduction}

Experimental measurements of the yield and elliptic flow of light nuclei in heavy ion collisions have been carried out at the BNL Relativistic Heavy Ion Collider (RHIC)~\cite{Adler:2001uy,Abelev:2010rv,Agakishiev:2011ib,Yu:2017bxv} and the CERN Large Hadron Collider (LHC)~\cite{Adam:2015yta,Adam:2015vda,Acharya:2017dmc}. To understand these results, both the statistical model~\cite{Andronic:2010qu,Cleymans:2011pe} and the coalescence model~\cite{Chen:2012us,Chen:2013oba,Shah:2015oha,Sun:2015jta,Sun:2015ulc,Zhu:2015voa,Yin:2017qhg} have been used.  In particular, the elliptic flows of light nuclei, such as deuteron (${\rm d}$), triton (${\rm t}$), and helium-3 ($^3{\rm He}$), have been measured by the STAR Collaboration at RHIC for Au+Au collisions at a wide range of collision energies from $\sqrt{s_{NN}}=7.7$ GeV to $200$ GeV~\cite{Adamczyk:2016gfs}.  Using a blast-wave model with its parameters fitted to the proton transverse momentum spectrum and elliptic flow, it was shown in Ref.~\cite{Adamczyk:2016gfs} that simply replacing the proton mass by those of light nuclei, the model failed to describe the experimental data on the elliptic flows of these light nuclei.   Instead, the measured light nuclei elliptic flows were found to be consistent with the results obtained from the coalescence model that uses the kinetic freeze-out nucleons from a multiphase transport (AMPT) model~\cite{Lin:2004en}.  A similar conclusion was found in Ref.~\cite{Yin:2017qhg} using the coalescence model based on the nucleon phase-space distribution from an extended blast-wave model. Besides fixing its parameters by fitting to the measured proton transverse momentum spectrum and elliptic flow, this extended blast-wave model further takes into consideration of the larger in-plane than out-of-plane flow velocity due to the positive elliptic flow by introducing a space-momentum correlation in the nucleon phase-space distribution. Specifically, nucleons with large transverse momentum are assumed to be more spread in space when the in-plane components of their momenta are larger than their out-of-plane components.    Contrary to the conclusion obtained from heavy ion collisions at RHIC, the ALICE Collaboration at LHC has found that its measured deuteron spectra and elliptic flows in Pb+Pb collisions at $\sqrt{s_{NN}}=2.76$ TeV~\cite{Acharya:2017dmc} are more consistent with the blast-wave model than the naive coalescence model that assumes that 
the deuteron elliptic flow at certain transverse momentum is twice the proton elliptic flow at half the deuteron transverse momentum~\cite{Kolb:2004gi}.  In the present study, we extend the study of Ref.~\cite{Yin:2017qhg} for heavy ion collisions at RHIC to that at LHC to show that the coalescence model using the nucleon phase-space distribution from the extended blast-wave model can also describe the experimental data from the LHC.

The paper is organized as follows. In the next section, we describe in detail the extended blast-wave model.  Section~III gives a brief review of the coalescence model that has been extensively used in studying light nuclei production in heavy ion collisions.  We then show and discuss in Section~\ref{results} the results on deuteron and helium-3 transverse momentum spectra and elliptic flows at the centralities of 10-20\% and 30-40\% in Pb+Pb collisions at $\sqrt{s_{NN}}=2.76$ TeV. Finally, a summary is given in Section~\ref{summary}.

\section{The extended blast-wave model}\label{blast}

In the conventional blast-wave model for nucleon production in heavy ion collisions, the invariant momentum spectrum of nucleons emitted from a fireball produced in heavy ion collisions is given by~\cite{Yin:2017qhg}
\begin{eqnarray}
E\frac{d^3N}{d^3{\bf p}}=\int_{\Sigma^\mu}d^3\sigma_\mu p^\mu f(x,p),
\end{eqnarray}   
where $\Sigma^\mu$ is the hyper-surface of the fireball with a covariant normal vector $\sigma_\mu$ and $p^\mu$ is the four-momentum of the emitted nucleon.  The Lorentz-invariant thermal distribution of nucleons emitted from the hyper-surface is $f(x,p)=\frac{2\xi}{(2\pi)^3}\exp\{-p^\mu u_\mu/T_K^{}\}$, where $\xi$ is the fugacity, $u_\mu$ is the four-velocity of a fluid element on the hyper-surface, and $T_K$ is the kinetic freeze-out temperature of the fireball. 

In terms of the longitudinal and transverse flow rapidities $\eta=\frac{1}{2}\ln\frac{t+z}{t-z}$ and $\rho=\frac{1}{2}\ln\frac{1+|\bm\beta|}{1-|\bm\beta|}$ with $\bm\beta$ being the transverse flow velocity, and the azimuthal angles $\phi_p$ and $\phi_b$ of the nucleon transverse momentum and the transverse flow velocity with respect to the reaction plane, respectively, one has 
\begin{eqnarray}
p^\mu u_\mu&=&m_T\cosh\rho\cosh(\eta-y)\nonumber\\
&&-p_T\sinh\rho\cos(\phi_p-\phi_b),\nonumber\\
p^\mu d^3\sigma_\mu&=&\tau m_T\cosh(\eta-y)d\eta rdr d\phi.
\end{eqnarray}   
In the above, $m_T=\sqrt{m^2 + p_T^2}$ is the nucleon transverse mass with $m$ being its mass; $r$ and $\phi$ are, respectively, the radial and angular coordinates of the nucleon in the transverse plane with the origin taken to be the center of the fireball; and $\tau=\sqrt{t^2-z^2}$ is the proper time with $z$ being the nucleon coordinate along the beam direction. Assuming that all nucleons are emitted at the proper time $\tau_0$, the invariant nucleon momentum spectrum becomes  
\begin{widetext}
\begin{eqnarray}
\frac{d^3N}{p_Tdp_Tdyd\phi_p}=\frac{2\xi\tau_0}{(2\pi)^3}\int_{\Sigma^\mu} d\eta rdr d\phi m_T\cosh(\eta-y)\exp{\left[-\frac{m_T\cosh\rho\cosh(\eta-y)-p_T\sinh\rho\cos(\phi_p-\phi_b)}{T_K}\right]}.
\end{eqnarray}
\end{widetext}

Same as in Refs.~\cite{Oh:2009gx,Yin:2017qhg}, we parametrize the transverse flow velocity by
\begin{equation}
\bm{\beta}=\beta(r)\left[1+\varepsilon(p_T^{})\cos(2\phi_b)
\right] \hat{\bm{n}},
\label{beta}
\end{equation}
where $\hat{\bm{n}}$ is the unit vector in the direction of their transverse flow velocity $\bm{\beta}$, which is taken to be normal to the surface of the firball. The radial flow velocity $\beta(r)$ and $p_T^{}$-dependent coefficient $\varepsilon(p_T)$ are parameterized as $\beta_0r/R$ and $c_1^{} \exp(-p_T^{}/c_2^{})$, respectively, with $R$ being the average transverse radius of the fireball.  The space-momentum correlation of in plane ($|p_{Tx}|>|p_{Ty}|$) nucleons with momentum greater than $p_0$ is introduced by letting $R=R_0~e^{a(p_T-p_0)}$.  The nucleons are then assumed to be uniformly distributed inside a cylinder and their spatial distribution has an elliptic shape in the transverse plane, so the spatial region can be expressed as
\begin{equation}\label{radius}
r \le R \left[ 1 + s_2^{} \cos(2\phi) \right],
\end{equation}
with the spatial eccentricity $s_2^{} =\langle (x^2 - y^2)/(x^2 + y^2)\rangle$~\cite{Lin:2001zk} and $\phi$ being the azimuthal angle of the nucleon position vector ${\bf r}$ in the transverse plane. 

The phase space distribution of freeze-out nucleons in this extended blast-wave model is thus characterized by the ten parameters: $\xi$, $\tau_0$, $T_K$, $\beta_0$, $R_0$, $c_1$, $c_2$, $a$, $p_0$, and $s_2$.

\section{the coalescence model}\label{coalescence}

In the coalescence model for nuclei production~\cite{Mattiello:1996gq,Chen:2003qj,Chen:2003ava}, the production probability of a nucleus of $Z$ protons and $N$ neutrons from $A=Z+N$ nucleons is given by the overlap of the Wigner function $f_A({\bf x}_1^\prime, ... ,{\bf x}_Z^\prime,{\bf x}_1^\prime, ... ,{\bf x}_N^\prime; {\bf p}_1^\prime, ... ,{\bf p}_Z^\prime,{\bf p}_1^\prime, ... ,{\bf p}_N^\prime,t^\prime)$ of the nucleus with the phase-space distribution of $f_p({\bf x},{\bf p},t)$ of protons and $f_n({\bf x},{\bf p},t)$ of neutrons at kinetic freeze-out, that is
\begin{eqnarray}
\label{coal}
&&\frac{dN_A}{d^3 {\mathbf P}_A}=g_A\int \Pi_{i=1}^Zp_i^\mu d^3\sigma_{i\mu}\frac{d^3{\bf p}_i}{E_i}f_p({\bf x}_i, {\bf p}_i,t_i)\nonumber\\
&&\times\int \Pi_{j=1}^Np_j^\mu d^3\sigma_{j\mu}\frac{d^3{\bf p}_j}{E_j}f_n({\bf x}_j, {\bf p}_j,t_j)\nonumber\\
&&\times f_A({\bf x}_1^\prime, ... ,{\bf x}_Z^\prime,{\bf x}_1^\prime, ... ,{\bf x}_N^\prime; {\bf p}_1^\prime, ... ,{\bf p}_Z^\prime,{\bf p}_1^\prime, ... ,{\bf p}_N^\prime;t^\prime)\nonumber\\
&&\times \delta^{(3)}\left({\bf P}_A-\sum_{i=1}^Z{\bf p}_i-\sum_{j=1}^N{\bf p}_j\right),
\end{eqnarray}
where $g_A=(2J_A+1)/2^A$ is the statistical factor for $A$ nucleons of spin $1/2$ to form a nucleus of angular momentum $J_A$.  The coordinate ${\bf x}_i$ and momentum ${\bf p}_i$ are those of the $i$-th nucleon in the fireball frame. The corresponding coordinate ${\bf x}_i^\prime$ and momentum ${\bf p}_i^\prime$, which appear in the Wigner function, are obtained by Lorentz transforming to the rest frame of produced nucleus and then propagating earlier freeze-out nucleons freely to the time when the last nucleon in the nucleus freezes out.

For the Wigner function of a nucleus, it is obtained from the Wigner transform of its wave function, which is taken to be the product of those of a harmonic oscillator potential with the oscillator constant determined by fitting the empirical charge radius of the nucleus. For deuteron and helium-3 studied in the paper, their Wigner functions are~\cite{Song:2012cd}
\begin{eqnarray}
f_2(\boldsymbol\rho,{\bf p}_\rho)=8g_2\exp\left[-\frac{\boldsymbol\rho^2}{\sigma_\rho^2}-{\bf p}_\rho^2\sigma_\rho^2\right],
\label{two}
\end{eqnarray}
and 
\begin{eqnarray}
&&f_3(\boldsymbol\rho,\boldsymbol\lambda,{\bf p}_\rho,{\bf p}_\lambda)\nonumber\\
&&=8^2g_3\exp\left[-\frac{\boldsymbol\rho^2}{\sigma_\rho^2}-\frac{\boldsymbol\lambda^2}{\sigma_\lambda^2}-{\bf p}_\rho^2\sigma_\rho^2-{\bf p}_\lambda^2\sigma_\lambda^2\right],
\label{three}
\end{eqnarray}
respectively,
where
\begin{eqnarray}\label{rel}
&&\boldsymbol\rho=\frac{{\bf x}_1^\prime-{\bf x}_2^\prime}{\sqrt{2}},\quad{\bf p}_\rho=\frac{{\bf p}_1^\prime-{\bf p}_2^\prime}{\sqrt{2}},\\
&&{\boldsymbol\lambda}=\frac{{\bf x}_1^\prime+{\bf x}_2^\prime-2{\bf x}_3^\prime}{\sqrt{6}},\quad{\bf p}_\lambda=\frac{{\bf p}_1^\prime+{\bf p}_2^\prime-2{\bf p}_3^\prime}{\sqrt{6}}.
\end{eqnarray}
In the above, the same masses for proton and neutron have been assumed.

Information on the statistical factors and the values of the width parameters in the Wigner functions for ${\rm d}$ and $^3{\rm He}$ as well as the empirical values of their charge radii and the resulting oscillator constants can be found in Ref.~\cite{Yin:2017qhg}.

\section{Results}\label{results}

\begin{table*}
\centering
\caption{Values of blast-wave model parameters for Pb+Pb collisions at $\sqrt{s_{NN}}=2.76$ TeV.}
\label{parameters} 
\medskip      
\begin{tabular}{c|c|c|c|c|c|c|c|c|c|c}
\hline
Centrality (\%) & $\xi$ & $\tau_0~({\rm fm}/c)$ & $T_K~({\rm MeV}) $ & $\beta_0$ & $R_0~({\rm fm})$ & $c_1$ & $c_2~({\rm GeV}/c)$ & $s_2$ & $p_0~({\rm GeV}/c)$ & $a~({\rm GeV}/c)^{-1}$ \\
\hline\hline
10-20 & 5.5 & 13.5 & 120 & 0.84 & 17.0 & 0.09 & 4.6 & -0.07 & 0.9 & 0.05 \\
30-40 & 5.0 & 10.5 & 120 & 0.825 & 13.0 & 0.15 & 3.3 & -0.12 & 0.9 & 0.02 \\
\hline
\end{tabular}
\end{table*}

\begin{figure}[h]
\centerline{~~~~~~~~~~~~~~~~\includegraphics[width=11.5cm]{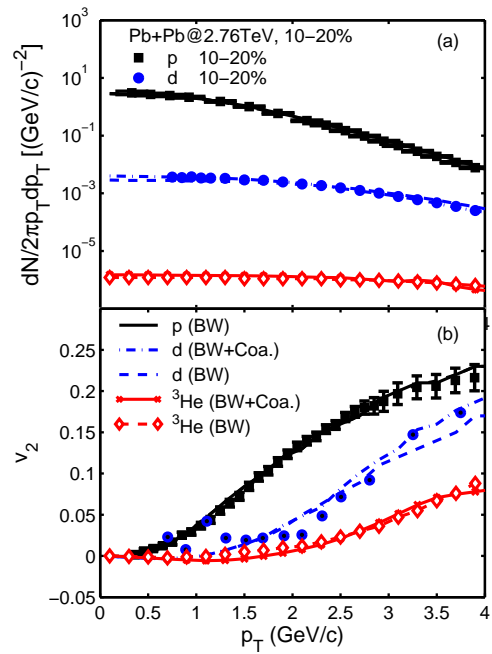}}
\caption{(Color online) Transverse momentum spectra (a) and elliptic flows (b) of midrapidity proton (p), deuteron (${\rm d}$) and helium-3 ($^3{\rm He}$) from the blast-wave model (BW) and the coalescence model (BW+Coa.) for Pb+Pb collisions at $\sqrt{s_{NN}}=2.76$ TeV and centrality of 10-20\%. Data for the transverse momentum spectra are taken from Ref.~\cite{Adam:2015kca} for proton and Ref.~\cite{Adam:2015vda} for deuteron, while those for elliptic flows are taken from Ref.~\cite{Abelev:2014pua} for proton and Ref.~\cite{Acharya:2017dmc} for deuteron.} 
\label{pt1020}
\end{figure}

Using nucleons from the extended blast-wave model described in Section II, we first fix the parameters of the model by fitting the measured proton transverse momentum spectrum and elliptic flow, shown by solid squares in Fig.~\ref{pt1020}, from the ALICE Collaboration for Pb+Pb Collisions at $\sqrt{s_{NN}}=2.76$ TeV and centrality of 10-20\%. The elliptic flow is calculated according to 
\begin{equation}\label{elliptic}
v_2=\left\langle \frac{p_x^2 - p_y^2}{p_x^2 + p_y^2}\right\rangle,
\end{equation}
where $p_x$ and $p_y$ are, respectively, the projections of the nucleon transverse momentum along the $x$ and $y$ axes in the transverse plane, which is perpendicular to the reaction plane .  These results are shown, respectively, in Fig.~\ref{pt1020}(a) and Fig.~\ref{pt1020}(b) by black solid lines, and they are obtained with values of the blast-wave parameters given in the first row of Table~\ref{parameters}. 

Although the proton transverse momentum spectrum and elliptic flow as well as the transverse momentum spectrum of deuteron are not affected by the parameters $p_0$ and $a$, the deuteron elliptic flow depends on their values.  In order to describe the measured deuteron elliptic flow (blue solid circles in Fig.~\ref{pt1020}(b)), it requires $p_0=0.9$ GeV/$c$ and $a=0.05~({\rm GeV}/c)^{-1}$, as shown by the blue dash-dotted line in Fig.~\ref{pt1020}(b).  Also shown in Fig.~\ref{pt1020} by blue dashed lines are the deuteron transverse momentum spectrum (a) and elliptic flow (b) obtained from the blast-wave model by simply replacing the proton mass with deuteron mass and using a deuteron fugacity of $\xi_d=14.0$.  The resulting deuteron transverse momentum spectrum is similar to that from the coalescence model and thus the experimental data.  However, the deuteron elliptic flow is smaller than that from the coalescence model when the momentum is larger than about 3 GeV/$c$ and is therefore below the measured values.  We have also calculated the transverse momentum spectrum and elliptic flow of helium-3 from the coalescence model. As shown by red solid lines in Fig.~\ref{pt1020}, they are very similar to those obtained from the blast-wave model using the mass of helium-3 and a helium-3 fugacity of $\xi_{^3{\rm He}}=15.5$, shown by red dashed lines. The results for the elliptic flow of helium-3 from the two models are, however, different at larger transverse momentum than that shown in the figure. 

\begin{figure}[h]
\centerline{~~~~~~~~~~~~~~~~\includegraphics[width=11.5cm]{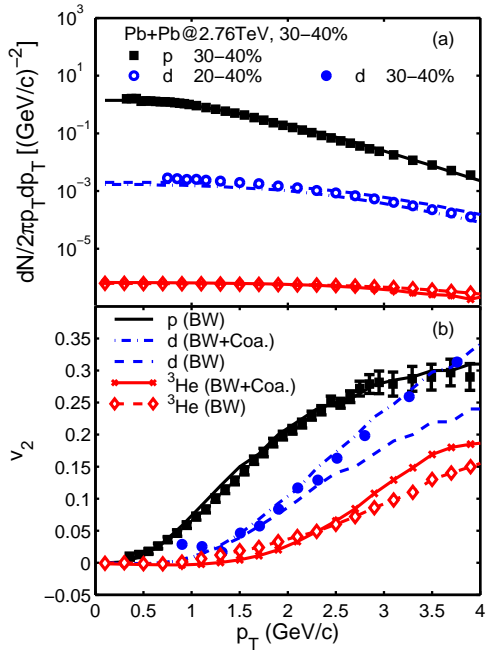}}
\caption{(Color online) Same as Fig.~\ref{pt1020} for centrality of 30-40\% except experimental data for the deuteron transverse momentum spectrum is at centrality 20-30\% and is shown by blue open circles.}
\label{pt3040}
\end{figure}

We have repeated the above calculations for the centrality of 30-40\% in Pb+Pb collisions at $\sqrt{s_{NN}}=2.76$ TeV using values of the blast-wave model parameters given in the second row of Table~\ref{parameters}.  The resulting transverse momentum spectra and elliptic flows of proton, deuteron, and helium-3 are shown in Fig~\ref{pt3040}(a) and Fig.~\ref{pt3040}(b), respectively.  It is seen that results from the coalescence model for the deuteron, shown by blue dash-dotted lines, reproduce very well the experimental data measured at the centrality of 20-40\% for spectrum (blue open circles) and 30-40\% for elliptic flow (blue solid circles).  Compared to the results from the blast-wave model by using the deuteron mass and fugacity $\xi_d=13.0$, the two models again agree in the deuteron transverse momentum spectrum but differ in the elliptic flow, with that from the blast-wave model visibly smaller for transverse momentum larger than about 2 GeV/$c$.  Deviation in the helium-3 elliptic flows from the coalescence model and the blast-wave model with a helium-3 fugacity of $\xi_{^3{\rm He}}=14.0$, shown by red solid and dashed lines, also becomes obvious for transverse momentum larger than about 2.5 GeV/$c$ with the coalescence model giving a larger value than the blast-wave model.  We note that the difference between the two models for the elliptic flows of deuteron and helium-3 at centrality 30-40\% is much larger than that at centrality 10-20\%, which suggest that the results from centrality at 30-40\% in Pb+Pb collisions at $\sqrt{s_{NN}}=2.76$ TeV can provide a stronger evidence to distinguish different models.

\section{summary}\label{summary}

Using an extended blast-wave model, which includes a space-momentum correlation in the phase-space distribution of high momentum nucleons, with its parameters fitted to measured proton transverse momentum spectrum and elliptic flow from Pb+Pb collisions at $\sqrt{s_{NN}}=2.76$ TeV for the two centralities of 10-20\% and 30-40\%, we have used the coalescence model to calculate the transverse momentum spectra and elliptic flows of deuteron and helium-3.  Our results for deuterons are seen to agree with the experimental data from the ALICE Collaboration. On the other hand, the deuteron elliptic flow obtained from the blast-wave model by using the deuteron mass fails to describe the data at large transverse momentum and is thus smaller than the results from the coalescence model.  For collisions at the centrality 30-40\%, a similar difference is found between the helium-3 elliptic flows at large transverse momentum obtained from the coalescence model and the blast-wave model using the helium-3 mass.  Our results thus show that the coalescence model using nucleons from the extended blast-wave model can describe the elliptic flow of deuterons measured in Pb+Pb collisions at LHC, as shown before for the elliptic flows of deuteron and helium-3 measured in Au+Au collisions at RHIC~\cite{Yin:2017qhg}. Therefore, studying light nuclei production provides the opportunity to probe the properties of the emission source of nucleons in relativistic heavy ion collisions, complimenting the study based on the Hanbury-Brown-Twiss (HBT) interferometry of identical particles emitted at freeze-out~\cite{Bertsch:1988db,Pratt:1990zq,Mrowczynski:1992gc}.

\section*{Acknowledgements}

The authors would like to thank Jurgen Schukraft for bringing the ALICE data on deuteron to their attention and for helpful communications. One of the authors (C.M.K.) is grateful to the Physics Department at Sichuan University for the warm hospitality during his visit 
when this work was carried out.  This work was supported in part by the NSFC of China under Grant no. 11205106, the US Department of Energy under Contract No. DE-SC0015266, and The Welch Foundation under Grant No. A-1358.

%\bibliography{references}

\end{document}